\documentclass[twocolumn,showpacs,prb,superscriptaddress,a4paper]{revtex4}
\usepackage[pdftex]{graphicx}
\usepackage{amsfonts}
\usepackage{amsmath}
\usepackage{amssymb}
\usepackage{color}

\newcommand{\ct}{\cite}
\newcommand{\bi}{\bibitem}
\newcommand{\be}{\begin{equation}}
\newcommand{\ee}{\end{equation}}
\newcommand{\ba}{\begin{eqnarray}}
\newcommand{\ea}{\end{eqnarray}}
\newcommand{\al}{\alpha}
\newcommand{\de}{\delta}

\newcommand{\si}{\sigma}
\newcommand{\om}{\omega}
\newcommand{\non}{\nonumber}

\newcommand{\bra}[1]{\langle #1|}
\newcommand{\ket}[1]{|#1\rangle}

\begin{document}

\title{Dynamical localization in a chain of hard core bosons under a 
periodic driving}
\author{Tanay Nag}
\affiliation{Indian Institute of Technology Kanpur, Kanpur 208016, India}
\author{Sthitadhi Roy}
\affiliation{Max-Planck-Institut f\"ur Physik komplexer Systeme, N\"othnitzer 
Strasse 38, 01187 Dresden, Germany }
\author{Amit Dutta}
\affiliation{Indian Institute of Technology Kanpur, Kanpur 208016, India}
\author{Diptiman Sen}
\affiliation{Centre for High Energy Physics, Indian Institute of Science, 
Bangalore 560012, India}

\begin{abstract}
We study the dynamics of a one-dimensional lattice model of hard core 
bosons which is initially in a superfluid phase with a current being induced 
by applying a twist at the boundary. Subsequently, the twist is removed and 
the system is subjected to periodic $\de$-function kicks in the staggered 
on-site potential. We present analytical expressions for the current
and work done in the limit of an infinite number of kicks. Using these,
we show that the current (work done) exhibit a number of dips (peaks) as a 
function of the driving frequency and eventually saturates to zero (a finite 
value) in the limit of large frequency. The vanishing of the current (and the
saturation of the work done) can be attributed to a dynamic localization of 
the hard core bosons occurring as a consequence of the periodic driving. 
Remarkably, we show that for some specific values of the 
driving amplitude, the localization occurs for any value of the driving 
frequency. Moreover, starting from a half-filled lattice of 
hard core bosons with the particles localized in the central region, we show 
that the spreading of the particles occurs in a light-cone-like region with 
a group velocity that vanishes when the system is dynamically localized.
\end{abstract}

\maketitle

\section{Introduction}

Periodically driven closed quantum systems have been studied extensively in 
recent years from the viewpoint of quenching dynamics as well as quantum 
information theory. Some of these systems show dynamical localization (DL) 
where the energy of the system never exceeds a maximum bound. Systems showing 
the signature of DL include driven two-level systems~\ct{grossmann91}, 
classical and quantum kicked rotors~\ct{chirikov81,ammann98} and the 
Kapitza pendulum~\ct{kapitza51}. In parallel, there have been several 
studies of many-body localization transition
which have indicated that disordered interacting systems can behave 
non-ergodically~\ct{basko06,pal10,polkovnikov13}. Given the recent interest in
quenching dynamics of quantum systems~\ct{dutta10,polkovnikov11,dziarmaga10} 
driven across a quantum critical point (QCP)~\ct{sachdev99,chakrabarti96}, the 
dynamics of those systems under a periodic modulation of the field has also 
been investigated~\ct{mukherjee09,das10}; the connection between 
thermalization and many-body localization has also been 
explored \ct{canovi12}. In particular, it has been observed 
that when a quantum many-body system, specifically, an Ising chain in a 
transverse magnetic field, is periodically driven across a QCP
there is a synchronization to a ``periodic" steady state~\ct{russomanno12}.

In this work, we study the dynamics of a chain of hard core bosons (HCBs) 
which is subjected to a periodic kick in the staggered on-site potential.
We address the issue of DL within the framework of 
Floquet theory applicable to a time-periodic Hamiltonian~\ct{shirley65}. 
Low-dimensional bosonic systems have been realized experimentally by trapping 
ultracold atoms in optical lattices~\ct{greiner01,bloch08} and the quantum 
phase transition from a superfluid (SF) to a Mott insulator (MI) phase has 
been observed in three dimensions~\ct{greiner02} as well as in one 
dimension~\ct{stoferle04}. The HCB system has also been realized 
experimentally in optical lattices~\ct{parades04,kinoshita04}. Following these
experimental realizations, there have been numerous analytical studies of 
these systems in recent years; for a review see Ref.~\onlinecite{cazalilla11}. 
The integrability of a HCB chain (and its continuum version known as 
the Tonks-Girardeau gas~\ct{tonks36}) has been exploited extensively, for 
instance, to investigate the surviving current when the HCB chain is quenched 
from the SF to the MI phase~\ct{klich07}, to study the quench dynamics when 
the system is released from a trap~\ct{collura13}, to analyze the origin of 
superfluidity out of equilibrium~\ct{rossini13}, and to explore the DL of 
bosons in an optical lattice~\ct{horstmann07}.

The paper is organized as follows. In Sec.~\ref{sec_floquet}, we introduce the
model, the initial state of the system (which carries a non-zero current), 
and the periodic driving scheme. 
We explicitly derive the Floquet operator and its eigenvalues for a single 
$\de-$function kick of the staggered potential. In Sec. ~\ref{sec_infty}, we 
present analytical and numerical results for the current and work done in the 
asymptotic limit of 
an infinite number of kicks. We analyze these results to highlight the 
light-cone-like propagation of the particles and the phenomenon of dynamical 
localization which occurs for certain driving amplitudes and for large 
driving frequencies. We make some concluding remarks in Sec.~\ref{sec_conclu}.

\section{The model and the Floquet operator}
\label{sec_floquet}

The model we consider here is a chain of HCBs on 
a lattice at half-filling described by the Hamiltonian 
\be \mathcal{H} = -w\sum_l (b_l^{\dagger}b_{l+1} + b_{l+1}^\dagger b_l) 
\label{eq:ham} \ee
where $b_l$'s are bosonic operators satisfying the commutation relations 
$[b_l,b_{l'}^{\dagger}]=\de_{l,l'}$ and the hard core condition $(b_l)^2=
(b_l^{\dagger})^2=0$, and $w$ (assumed to be positive) is the hopping 
amplitude. Using the Jordan-Wigner transformation from HCBs to 
spinless fermions~\ct{lieb61}, the Hamiltonian in
\eqref{eq:ham} can be mapped to a system of non-interacting fermions, 
which, in momentum space, gets decoupled into $2 \times 2$ Hamiltonians in 
terms of the momenta $k$ and $k+\pi$, where $-\pi/2 \le k \le \pi/2$.
Using the basis vector $\ket{k} =\left( \begin{array}{cc} 
1~~ 0 \end{array} \right)^T$ and $\ket{k +\pi} =\left( \begin{array}{cc} 
0~~ 1 \end{array} \right)^T$, one can rewrite the $2 \times 2$ Hamiltonians
as
\be \mathcal{H}_k ~=~ -2w\cos k ~\si^z, \label{eq:hamk} \ee
where $\si$'s denote the Pauli matrices. At half-filling, all the $k$-values 
from $-\pi/2$ to +$\pi/2$ are filled; the ground state for every $k$ mode
is the pseudo-spin up state of the operator $\si^z$ denoted by 
$\left( \begin{array}{cc}
1~~ 0 \end{array} \right)^T$.

When a staggered on-site potential (in real space) of the form 
$V\sum_l (-1)^l b_l^{\dagger}b_l$ is added to the Hamiltonian in 
\eqref{eq:ham}, a coupling is generated between the modes with momenta 
$k$ and $k+\pi$. Consequently, an energy gap opens up at $k=\pm\pi /2$, 
hence, the system is in the Mott insulator phase for any finite value of $V$. 
There is a quantum phase transition separating the gapped MI phase from the 
gapless SF phase when $V \to 0$. 

We now consider a boosted Hamiltonian, $\mathcal{H}_{\nu}=-
w\sum_l (b_l^{\dagger}b_{l+1} e^{-i\nu} + b_{l+1}^\dagger b_l e^{i\nu})$,
whose ground state has a non-zero current.
[We call $\nu$ a boost because it effectively shifts
the momentum from $k$ to $k-\nu$. Assuming periodic boundary conditions,
one can perform certain phase transformations on the $b_l$ to remove $\nu$ 
from each of the terms in the Hamiltonian except for the last term which hops 
from site $L$ to site 1; the phase of this hopping amplitude then becomes 
$L\nu$, where $L$ is the number of sites. Hence,
$\nu$ also describes a twist in the boundary condition]. With the twist, the 
ground state for the modes between $-\pi/2 < k <-\pi/2 + \nu$ is given by
$\left( \begin{array}{cc} 0~~ 1 \end{array} \right)^T$ (i.e., the $k+\pi$ 
modes are occupied), while for the modes between $-\pi/2+\nu < k <\pi/2$ the 
ground state is $\left( \begin{array}{cc} 
1~~ 0 \end{array} \right)^T$. We define the current operator $\hat{j}=-(1/L) 
(\partial H_{\nu}/\partial \nu)_{\nu=0} = (iw/L)\sum_l(b_{l+1}^{\dagger}b_l - 
b_l^\dagger b_{l+1})$, which takes the form $\hat{j}_k= (2 w/L)\sin k \si^z$ 
in the space of momenta $(k,k+\pi)$. In the limit $L \to \infty$, we find the 
initial current to be $j= (2w/\pi) \sin{\nu}$. (We will set 
$w=1$ below).

We will now remove the twist (at $t=0$) and study what happens to the initial 
current carrying state when a periodic perturbation is applied (that starts 
at $t=T$ with a period $T$) to the staggered potential. More specifically, we 
will focus on the situation when the Hamiltonian in \eqref{eq:ham} is 
subjected to a periodic staggered on-site potential of the form of a Dirac 
$\de$-function kick of amplitude $\al$, applied at regular intervals of time 
denoted by $T$, 
\be V(t) ~=~ -\al ~\sum_{n=1}^\infty \de(t-nT). \label{eq:quench} \ee 

The main question that we will address here is whether the initial current 
generated by the twist survives in the asymptotic limit ($t \to \infty$)
under these periodic perturbations even when the HCB chain is always in the 
superfluid state (except for the $\de$-function kicks at $t=nT$). The 
interesting result we would like to emphasize at the outset is that following 
an infinite number of kicks ($n \to \infty$), the current vanishes in the 
limit of large driving frequency $\om_0 = 2\pi/T$, while the excess energy 
energy saturates to a non-zero finite value. As will be discussed below,
the vanishing of the current can be attributed to a DL 
due to a decoherence which leads to a mixed density matrix at large times;
we will see that the probability of finding a boson at any site becomes 
equal $1/2$ for $\om_0 \to \infty$, for all values of $\al$. At the same 
time, the work done, $W_d$, saturates to a finite value. We will also show 
that for values of the kick amplitude $\al$ for which $\cos \al =0$, the 
current vanishes for $n \to \infty$ for all values of $\om_0$. 



We now recall the Floquet theory for a generic time-periodic Hamiltonian, 
$H\left(t\right)=H\left(t+T\right)$. One can construct a Floquet operator 
${\cal F}=\mathcal{T}e^{-i\int_0^T H(t)dt}$, where $\mathcal{T}$ denotes
time-ordering. The solution of the Schr\"odinger equation for the $j$-th 
state in the Floquet basis ($\ket{\Phi_j(t)}$ which are eigenstates of 
${\cal F}$) can be written in the form $\ket{\Psi_j(t)}=e^{-i\mu_j t}
\ket{\Phi_j(t)}$. The states $\ket{\Phi_j(t)}$'s 
are time periodic ($\ket{\Phi_j(t)}=\ket{\Phi_j(t+T)}$) and $e^{-i\mu_j T}$
are the corresponding eigenvalues of ${\cal F}$; the $\mu_j$'s are called 
Floquet quasi-energies. To study the dynamics of the Hamiltonian in 
\eqref{eq:ham} under the periodic kicks, we note that the Floquet 
operator in momentum space is given by
\be \mathcal{F}_k=\exp(-i P_k)\exp(-i \mathcal{H}_k T), \label{eq:flo} \ee 
where $P_k=-\al \si^x$ and ${\mathcal H}_k=-2\cos k \si^z$. The first term in 
\eqref{eq:flo} represents time evolution due to the $\de$-function kick 
at time $t=T$ while the second term denotes the time 
evolution of the system dictated by the Hamiltonian in
\eqref{eq:hamk} for an interval of time $T$. Looking at the form of the 
Floquet operator, one immediately finds some specific values of $\al$ given by 
$\al =m\pi$, where $m=0,1,2,\cdots$, for which the $\de$-function 
kicks do not affect the temporal evolution of the HCB chain; the ground state 
remains frozen in its initial state.

The expression for the Floquet operator for a single kick can be obtained 
exactly \ct{sauer12}:
\be \mathcal{F}_k = \left[ \begin{array}{cc} 
a & b \\
-b^* & a^* \end{array} \right], \label{eq:specf} \ee
where $a= \cos \al \cos(2T \cos k ) + i \cos\al \sin(2T \cos k)$, 
$b= \sin \al \sin(2T \cos k) + i \sin\al \cos(2T \cos k)$. The eigenvalues 
of the operator in (\ref{eq:specf}) are ${\rm e}^{i \mu_k^{\pm}T}$, where 
\be \mu_k^{\pm}T ~=~ \pm \arccos [\cos \al \cos(2T \cos k )],
\label{eq_quasi_energy}
 \ee
lie in the range $[-\pi,\pi]$. The Floquet quasi-states 
$\ket{\Phi_k^{\pm}}$ are given by the eigenstates of $\mathcal{F}_k$.
It is clear from the structure of \eqref{eq:specf} that it is sufficient
to consider values of $\al$ lying in the range $[0,\pi]$. Further, 
$\mathcal{F}_k$ for $\al=0$ and $\al=\pi$ only differ by a minus sign;
hence all the physical properties of the system are the same at these two
values of $\al$ as we will show below.

\begin{figure}[htb]
\begin{center}
\includegraphics[width=0.95\columnwidth]{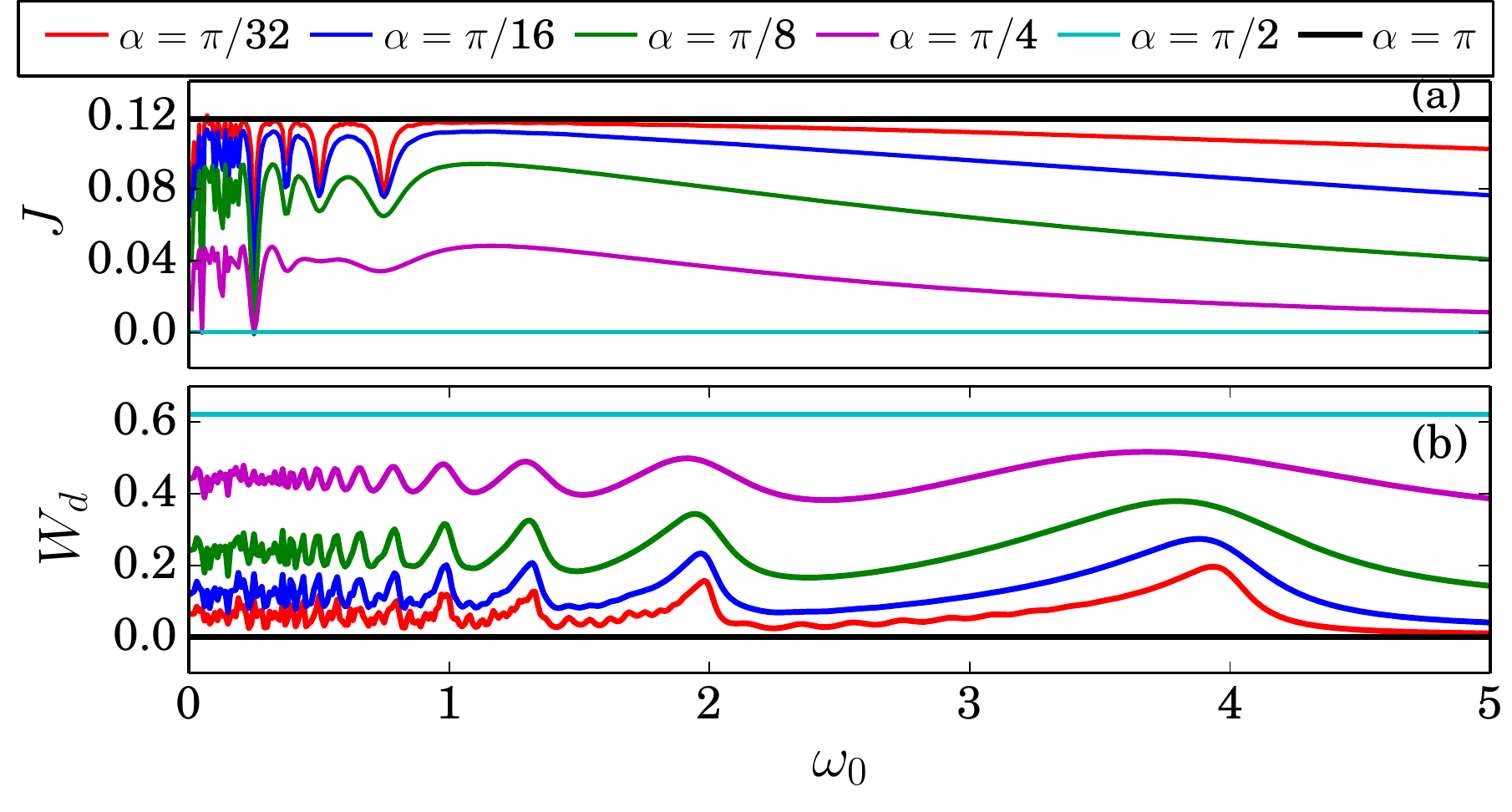}
\end{center}
\caption{(Color online) Plots of (a) current $J$ and (b) work 
done $W_d$ as functions of $\om_0$ for small values of $\om_0$ and 
several values of $\al$, with $L=100$ and $\nu=0.2$.
$W_d$ has peaks at some specific values of $\om_0$ given by $4/n$, where 
$n$ is an integer, which are the quasi-degeneracy points of the Floquet 
spectrum. The positions of the dips in $J$ are different from the peak 
positions in $W_d$ as explained in the text.} \label{J_WF} \end{figure}

\begin{figure}[htb]
\begin{center}
\includegraphics[width=0.95\columnwidth]{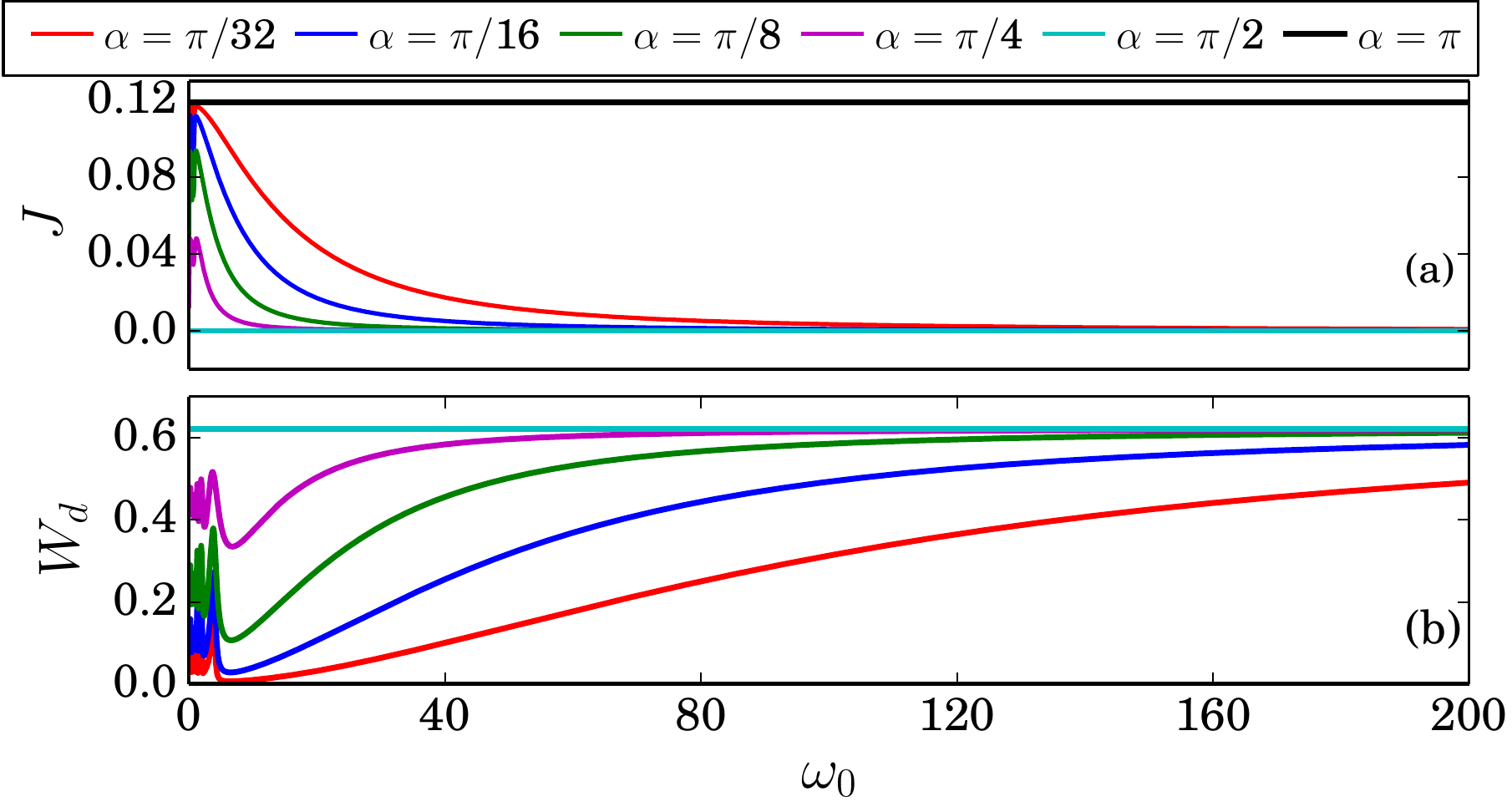}
\label{J_WF_1}
\end{center}
\caption{(Color online) Plots of (a) current $J$ and (b) work done $W_d$ as 
functions of $\om_0$ for large values of $\om_0$ and several values of $\al$, 
with $L=100$ and $\nu=0.2$. $J$ ($W_d$) stays at a higher (lower) value 
for small values of $\al$. 
$J$ and $W_d$ asymptotically saturate to zero and a finite value respectively.
For the special value $\al=\pi$, $J$ sticks to the initial value as $\om_0$ is
varied, while for $\al=\pi/2$, $J$ always stays at zero.} \label{J_W} 
\end{figure}

Under the periodic driving, the time evolved state at time 
$t=nT$ can be obtained by $n$ applications of the Floquet operator, namely, 
$\ket{\Psi_k(nT)}$ $=c_k^+ \mathrm{e}^{-i\mu_k^+ \,nT} \ket{\Phi_k^+} + c_k^- 
\mathrm{e}^{-i\mu_k^- \,nT} \ket{\Phi_k^-}$, where $c_k^{\pm}=\langle 
\Phi_k^\pm \ket{\Psi_k(0)}$, with $\ket{\Psi_k(0)}$ being the ground state of 
the Hamiltonian in \eqref{eq:hamk}. We can then compute the current $J(nT)=
\sum_k J_k (nT) \equiv \sum_k \bra{\Psi_k(nT)} \hat{j}_k \ket{\Psi_k(nT)}$,
and the work done $W_d = (1/L) \sum_k W_k \equiv (1/L) \sum_k [e_k(nT)-
e_k(0)]$, where $e_k(nT)$ is the energy of the $k$-th mode measured after $n$ 
kicks, given by $e_k(nT)= \bra{\Psi_k(nT)} \mathcal{H}_k \ket{\Psi_k(nT)} =-2
\bra{\Psi_k(nT)} \si^z \ket{\Psi_k(nT)} \cos k$, and $e_k(0) = -2 
\bra{\Psi_k(0)} \si^z \ket{\Psi_k(0)} \cos k$ is the initial ground state 
energy. 

\section{The $n\to \infty$ limit: results and implications}
\label{sec_infty}

We now consider the limit $n \to \infty$ when $\bra{\Psi_k(nT)} \si^z
\ket{\Psi_k(nT)} = \sum_{m=\pm}|c_k^m|^2 \bra{\Phi_k^m} \si^z \ket{\Phi_k^m}$,
where we have dropped rapidly oscillating cross-terms (with coefficients 
$c_k^{+*}c_k^-$ and $c_k^{-*}c_k^+$) which decay to zero in the limit 
$t \to \infty$ when integrated over a large number of momenta modes.
Given the initial ground state with a twist, we find that 
\ba \sum_{m=\pm} && \hspace*{-.5cm} |c_k^m|^2 ~\bra{\Phi_k^m} \si^z 
\ket{\Phi_k^m} \non \\ 
&=& -~ f(k) ~~{\rm for}~ -\pi/2 \le k \le - \pi/2 + \nu, \non \\
&=& f(k) ~~{\rm for}~ -\pi/2 + \nu \le k \le \pi/2, \non \\
{\rm where}~~ f(k) &=& \frac{\cos^2 \al ~\sin^2 (2T \cos k)}{\sin^2 \al ~+~ 
\cos^2 \al ~ \sin^2 (2T \cos k)} . \label{phimk} \ea
These expressions imply that the properties of the system will
remain the same if we change $\al \to \al + \pi$ or $\pi - \al$.

We will eventually be interested in the thermodynamic limit $L \to \infty$
where we replace $(2\pi/L) \sum_k \to \int dk$. We then obtain the following 
expressions for the current and work as $n, L \to \infty$,
\ba J (\infty) &=& \frac{1}{\pi} \int_{-\pi/2}^{\pi/2} dk \sum_{m=\pm}
|c_k^m|^2 \bra{\Phi_k^m} \si^z \ket{\Phi_k^m} \sin k \non \\
&=& -~\frac{2}{\pi} \int_{-\pi/2}^{-\pi/2 +\nu} dk ~f(k) ~\sin k, \non \\
W_d (\infty) &=& \frac{2}{\pi} \cos \nu \non \\
&& - \frac{1}{\pi} ~\int_{-\pi/2}^{\pi/2} dk \sum_{m=\pm} |c_k^m|^2 
\bra{\Phi_k^m} \si^z \ket{\Phi_k^m} \cos k \non \\
&=& \frac{2}{\pi} \cos \nu ~-~ \frac{1}{\pi} \int_{-\pi/2+\nu}^{\pi/2 +\nu} 
dk ~f(k) ~\cos k, \label{jwd} \ea
where the first term in the last two equations comes from $-(1/2\pi) 
\int_{-\pi/2}^{\pi/2} dk \bra{\Psi_k(0)} \si^z \ket{\Psi_k(0)} \cos k 
= (2/\pi) \cos \nu$. We will denote $J(\infty)$ and $W_d (\infty)$ by
$J$ and $W_d$ below.

The expressions in \eqref{phimk} vanish in two cases:
(i) $T \to 0$, i.e., the driving frequency $\om_0 \to \infty$, while
$\al$ may take any value, and 
(ii) $\cos \al = 0$, i.e., $\al = (m+1/2)\pi$, while $T$ may take any
value. In these two cases, we obtain $J = 0$ and $W_d = (2/\pi) \cos \nu$.

The neglect of the cross-terms in the limit $n \to \infty$ as discussed 
earlier implies that we have a decohered density matrix. The special feature 
of the two cases $\om_0 \to \infty$ (and any $\al$) and $\cos \al = 0$ 
(and any $\om_0$) is that $|c_k^+|^2 = 
|c_k^-|^2 =1/2$ for all $k$; namely, the density matrix is given by $1/2$ 
times the identity matrix in the space of momenta $(k,k+\pi)$ for all $k$.
Since the density matrix is proportional to the identity, it is invariant
under all unitary transformations. In particular, we can transform to the 
position basis and conclude that the system is described by a mixed density 
matrix in which the probability of finding a boson at any site is equal 
to $1/2$. This corresponds to a completely localized state; this is like a 
classical state in which the bosons have a probability of $1/2$ of being at 
each site. This explains the vanishing of the current and the saturation of 
the work done in these two cases.

Using Eq.~\eqref{phimk} we can evaluate the leading order behaviors of
the quantities in \eqref{jwd} in various limits. In the limit 
$\om_0 = 2\pi/T \to \infty$, we find that
\be J \to \frac{32\pi}{3} \frac{\cot^2 \al ~\sin^3 \nu}{\om_0^2} ~~
{\rm and}~~ W_d \to \frac{2}{\pi} \cos \nu. \label{JWinf} \ee
Two other limits are of interest. For $\om_0 \to 0$, we find that
$J \to (2/\pi) \sin \nu (1 - |\sin \al|)$, while for
$\nu \to 0$, we find $J \to (32\pi/3) \nu^3 \cot^2 \al/\om_0^2$. The 
latter behavior has been called the $\nu^3$ law in Ref.~\onlinecite{klich07}.

Next we investigate the current $J$ and work done $W_d$ as functions of 
$\om_0$ for a wide range of $\om_0$, with different values of $\al$.
An examination of Figs.~\ref{J_WF} and \ref{J_W} shows three distinct 
regions where the current and work done show three different behaviors. (i) 
For smaller values of $\om_0$, $J$ shows dips at some specific values of 
$\om_0$, while $W_d$ exhibits peaks at $\om_0$ which are 
different from the positions of the dips in the current. (ii) In an 
intermediate region of frequency, $J$ decreases monotonically with increasing 
$\al$ up to $\al < \pi/2$ while $W_d$ increases in a similar fashion. (iii) 
Both quantities saturate asymptotically at some specific values in the 
large frequency limit. 

\begin{figure}
\begin{center}
\includegraphics[width=\columnwidth]{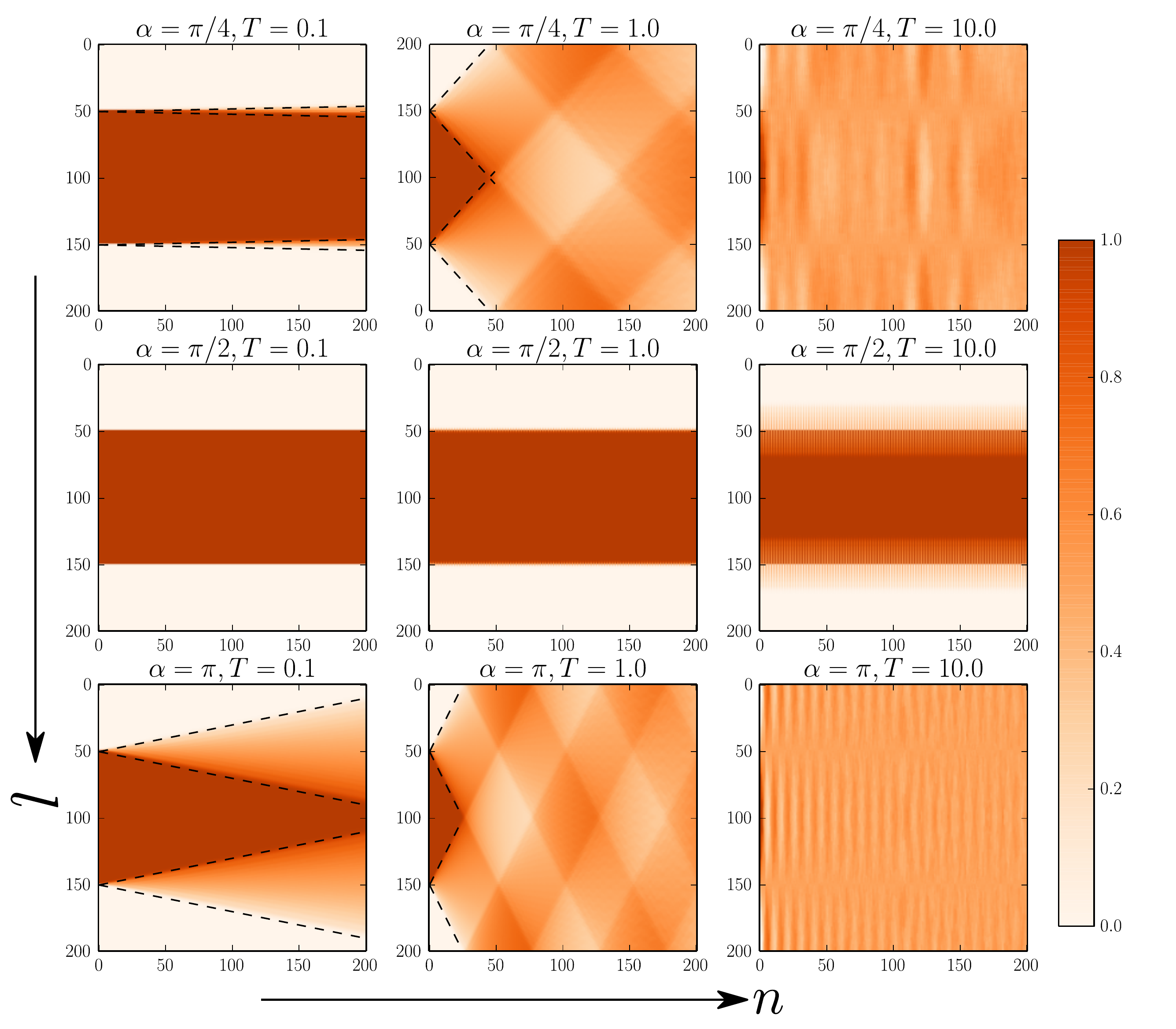}
\caption{(Color online) Pictures showing the density of particles in a 
200-site system as a function of the stroboscopic time $t=nT$ (on the $x$-axis)
and the location $l$ (on the $y$-axis) for various values of $T$ and $\al$. 
See text for details.} \label{fig_dyn_loc} \end{center} \end{figure}

We now discuss the positions of the peaks in $W_d$ and dips in $J$ in the 
small $\om_0$ regime as shown in Figs.~\ref{J_WF} (a) and \ref{J_WF} (b)
obtained through numerical studies of Eq.~(\ref{phimk}).
We will argue that the positions of the peaks in $W_d$ are related to the 
quasi-degeneracy of the Floquet spectrum near $k=0$. 
Since $W_k$ is 
proportional to $\cos k$, $W_d$ receives its largest contribution from the 
region near $k=0$. For small values of $\al$, the positions of the maxima in 
$W_d$ are therefore determined by the condition $2T (\cos k)|_{k=0} = m\pi$, 
i.e., $\om_0= 4/m$ where $m$ is an integer. Indeed we see that $W_d$ has 
peaks around $\om_0 = 4,~2,~1.3,\cdots$ in Fig.~\ref{J_WF} (b).
We now turn to the dips in $J$. For small values of $\nu$, we see from 
Eq.~\eqref{jwd} that the integral expression for $J$ goes over a small range 
from $-\pi/2$ to $-\pi/2 + \nu$. The integrand $f(k) \sin k$ vanishes at the 
lower limit $k=-\pi/2$; it also vanishes at the upper limit if $2T \cos (-\pi/2
+ \nu) = (4\pi/\om_0) \sin \nu = m\pi$, where $m$ is an integer. We therefore 
expect that the entire integral will show a dip as a function of $\om_0$ if 
$\om_0 = (4 \sin \nu)/m$. For $\nu = 0.2$, we expect $J$ to show dips around
$\om_0 = 0.8,~0.4,~0.26,\cdots$ as shown in Fig.~\ref{J_WF} (a). For large 
$\om_0$, $J$ ($W_d$) asymptotically saturate to zero ($(2/\pi) \cos \nu$); 
see Figs.~\ref{J_W} (a) and \ref{J_W} (b) where we show 
the variation of $J$ and $W$ for the entire range of $\om_0$.
We also see that $J$ approaches zero for smaller values of $\om_0$ as $\al$ 
increases; this is in accordance with Eq.~\eqref{JWinf} since $\cot \al$
decreases as $\al$ increases from zero to $\pi/2$.
 
Finally, we summarize our observations on the dependences of $J$ and $W_d$ on 
$\al$. (i) $J$ and $W_d$ remain at the constant values $(2/\pi) \sin \nu$ and 
zero for all $\om_0$ for the special values $\al =m\pi$. (ii) $J$ ($W_d$) 
remains at zero ($(2/\pi) \cos \nu$) for any $\om_0$ for $\al=(m+1/2)\pi$.
In this case, the Floquet quasi-states $\ket{\Phi_k^{\pm}}$ have zero 
expectation values for the matrix $\si^z$ appearing in the expression for the 
current. (iii) The magnitude of $J$ ($W_d$) decreases (increases) as 
$\al$ increases from zero to $\pi/2$. 

The DL which occurs in either of the 
limits $T \to 0$ or $\al = \pi/2$ is illustrated in Fig.~\ref{fig_dyn_loc}.
The figures show the density of particles in a 200-site system as a function 
of the time $t=nT$ (along the $x$-axis) and the location $l$ (along the
$y$-axis) for various values of $T$ and $\al$. The initial
state at $t=0$ is one in which sites 51 to 150 have one particle each
(shown by dark regions) and the remaining sites are empty (shown by light
regions). As $t$ increases, the particles spread out with group velocities
given by $v^\pm_k = d\mu^\pm_k/dk$. The spreading occurs in light-cone-like
regions whose slopes $dl/dt=(1/T)dl/dn$ are given by the maximum value of 
$|v^\pm_k|$ as a function of $k$; these are shown by the black dashed lines. 
It can be shown from Eq.~(\ref{eq_quasi_energy}) that the maximum velocity 
goes to zero as either $T \to 0$ or $\al \to \pi/2$. (For instance, if
$\al = \pi/2$, we find that $\mu_k^\pm = \pm \pi /(2T)$, so that $v_k^\pm
= 0$ for all $k$). This clearly demonstrates the
DL. While light-cone-like effects have been studied 
following a quantum quench both theoretically~\cite{lightcone} and 
experimentally~\cite{cheneau}, our work appears to be the first to study 
this in the context of periodic driving. (We remark that the ripples 
appearing in Fig.~\ref{fig_dyn_loc} in the panel for $\al=\pi/2, ~T=10.0$ are 
finite size effects).


\section{Conclusions}
\label{sec_conclu}

To summarize, we have explored the consequences of applying periodic 
$\de$-function kicks in the staggered on-site potential on the current 
carrying ground state of a HCB chain. 
In the long time limit ($n \to \infty$), there is an onset of DL if either 
the frequency of driving is large or the driving amplitude takes some 
particular values. We conclude with the remark that the DL occurring as a 
result of the periodic driving is not special to the one-dimensional model of 
hard core bosons discussed here; it can also be shown to occur in models of 
non-interacting fermions on a variety of higher-dimensional lattices (such
as square and cubic lattices) with a periodic driving of a staggered on-site 
potential. In the future it may be interesting to study the effect of 
interactions between fermions on DL.

We thank Achilleas Lazarides, G. E. Santoro and A. Russomanno for discussions,
and Shraddha Sharma and Abhiram Soori for critical comments. For financial 
support, D.S. thanks DST, India for Project No. SR/S2/JCB-44/2010.

\end{document}